# A Note on Extending Taylor's Power Law for Characterizing Human Microbial Communities: Inspiration from Comparative Studies on the Distribution Patterns of Insects and Galaxies, and as a Case Study for Medical Ecology and Personalized Medicine

Zhanshan (Sam) Ma
Computational Biology and Medical Ecology Laboratory
State Key Laboratory of Genetic Resources and Evolution
Kunming Institute of Zoology
Chinese Academy of Sciences, P. R. China
ma@vandals.uidaho.edu

**ABSTRACT**—Many natural patterns, such as the distributions of blood particles in a blood sample, proteins on cell surfaces, biological populations in their habitat, galaxies in the universe, the sequence of human genes, and the fitness in evolutionary computing, have been found to follow power law (*e.g.*, Kendal 2004, Illian et al. 2008, Venter 2001, Ma 2012). Taylor's power law (Taylor 1961: *Nature*, vol. 189:732–735.) is well recognized as one of the fundamental models in population ecology, thanks to its wide applicability in describing the spatial distribution patterns of biological populations. A fundamental property of biological populations, which Taylor's power law reveals, is the near universal heterogeneity (also known as aggregation or non-randomness) of population abundance distribution in habitat. Obviously, the heterogeneity also exists at the community level, where not only the distributions of population abundances but also the proportions of the species composition in the community are often heterogeneous or non-random. Indeed, community heterogeneity is simply a reciprocal term of community evenness, the major dimension what community diversity tries to measure. Nevertheless, existing community diversity indexes such as Shannon index and Simpson index can only measure "local" or "static" diversity in the sense that they are computed for each habitat at a specific time point, but the indexes alone do not reflect the diversity changes across habitats or over time. This inadequacy is particularly problematic if the research objective is to study the dynamics of community diversity, which is critical for understanding the possible mechanisms that maintain the community diversity and stability. In this note, I propose to extend the application scope of Taylor's power law to the studies of human microbial communities, specifically, the community heterogeneity at both population (single species) and community (multiple species) levels. I further suggested that population dispersion models such as Taylor (1980, *Nature*: 286, 53-55), which is known to generate population distribution patterns consistent with the power law, should also be very useful for analyzing the distribution patterns of human microbes within the human body. Finally, I suggest that the parameters of the power law model built at community level, especially when associated with community metadata (or environmental covariates) and when time series data from longitudinal studies are utilized, can reveal important dynamic properties of human microbiome such as community stability, which can be invaluable in investigating etiology of some diseases associated with human microbial communities. Overall, I hope that the approach to human microbial community with the power law offers an example that ecological theories can play an important role in the emerging *medical ecology*, which aims at studying the ecology of human microbiome and its implications to human diseases and health, as well as in personalized medicine.

## BACKGROUND AND JUSTIFICATIONS

Maurer (1999) pointed out two possibilities for the lack of generalizability in the study of community ecology. First, generalizations may be impossible due to the extreme complexity of ecological world. Second, patterns do not exist at the scale of the local community where the observation is made. Perhaps a third possibility can be added to this list: that the tool for detecting the patterns may be unsuitable at the research scale. While little may be done with respect to the first possibility, ecologists can strive to avoid the second and third possibilities by carefully choosing a right scale level and developing powerful approaches that are suitable for the chosen scale. In this note, I propose to extend the application scope of Taylor's power law (Taylor 1961, 1977, 1984, 2007) from population scale (level) to community scale (level) in order to obtain an appropriate and powerful quantitative approach for characterizing the microbial communities. The inspiration comes from a phenomenon in population ecology where Taylor's power law has been found to describe the spatial distribution patterns of biological populations in their habitat remarkably well. What Taylor's power law captures at the population scale is essentially the heterogeneity of population abundance distribution in space and/or time, and heterogeneity is certainly a fundamental property at community level (scale). Therefore, an interesting question is: can Taylor's power law play a similar role at the community level?

Taylor (1961) discovered that the Power Law model [Equation (1)] fits population spatial distribution data almost ubiquitously well from bacteria, plants, to mammals,

$$V = aM^b, \quad (1)$$

where $M$ and $V$ are population *mean* (density or abundance) and *variance*, respectively, and $a$ and $b$ are parameters. According to Taylor (1961, 1984): *b>1* corresponds to aggregated distribution (Fig. 1c) *b=1* to random distribution and (Fig. 1b) *b<1* to regular distribution (Fig. 1a).

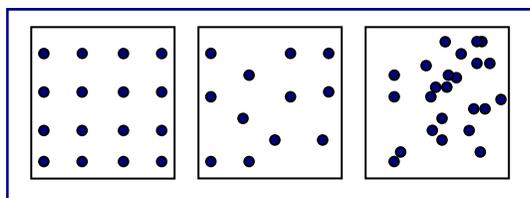

Fig. 1. Three Distribution Patterns of Biological Populations: (a): Regular, (b) Random, and (c) Aggregated.

Taylor's power-law accurately depicts the scaling relationship that describes the 'fat tails' of the distribution of abundance of nearly all organisms, and it has been found to hold at multiple scale levels of population ecology. Taylor's power law is one form of the power law that is found to underlie many natural processes such as the distribution of galaxies in the universe, blood cells in a blood sample. In addition, the three patterns displayed in Figure 1 are considered as typical patterns of *point process*, which is studied in a branch of spatial statistics, *i.e, point process statistics*.

The point process and power law have been studied in many fields of science; for example, in astronomy, both spatial random fields (*e.g.*, the density fields, the velocity fields) and spatial point patterns (*e.g.*, the position of galaxies) have been studied extensively (Illian et al. 2008). Astronomers had observed that matters in the universe is distributed in aggregation (clusters) on small scales, although, theoretically, the cosmological principle Einstein assumed in 1917 considered that the universe is isotropic and homogenous on large scale.

It is inspiring to note that Jerzy Neyman (1894-1981), a distinguished mathematical statistician from the University of California at Berkeley, studied the spatial distribution patterns of insects, bacteria, and galaxies, apparently with similar statistical approaches. Perhaps his study on galaxies distribution was inspired by his experience in studying bacteria and insects since his work in biology (Neyman 1939) was earlier than his work in astronomy (Neyman & Scott 1952, Neyman 1958). Even more inspiring is that the studies on the distribution patterns of insects and microbes in their habitat, as well as the studies on the distribution of galaxies in the universe, performed by Neyman and other scientists, evolved similarly in terms of the research approaches developed in astronomy and biology, respectively. In both astronomy and biology, three mathematical approaches with slightly different names were advanced; they are (*i*) frequency distribution—fitting the occurrence frequency data of subjects (galaxies or insects) per sampling unit to a probability distribution such as *Neyman distribution* (Neyman 1939) and *negative binomial distribution* (NBD); (*ii*) aggregation index (also known as dispersion index)—using some statistics to measure the aggregation (dispersion) degree of galaxies or insect populations; (iii) power law modeling.

While Taylor's power law [Equation (1)] describes the distribution patterns of biological populations, the form of power law for describing the distribution of galaxies takes a slightly different form:

$$1 + g(r) = \left(\frac{r_0}{r}\right)^s \quad (2)$$

where $g(r)$ is the so-termed pair correlation function, $s$ and $r_0$ are parameters. According to Illian et al. (2008), this power law model [Equation (2)] was proposed by Totsuji & Kihara (1969) to simplify Neyman et al's (Neyman &Scott 1952, 1958, Neyman et al. 1953) computation and to relieve the statistical difficulties Neyman et al (1953) encountered. Neyman et al. (1953) statistical difficulty was that they obtained different model parameters for different galaxy catalogues (files of coordinates of galaxies); therefore, their model was a probability distribution model that is a superposition of groups of galaxies of varying size, including super clusters. In fact, in biology, the discovery and application of Taylor's power law played a similar role in the study of the distribution of biological populations, especially in the study of insects (Taylor 1961, 1984). In entomology, from 1920's to 1950's, several probability distributions including negative binomial, Neyman, Thomas Double Poisson, Polya distributions, had been developed to fit the frequency data of insect population distribution. A practical problem with the probability distribution approach in entomology was that for each data set of population distribution, there is a set of parameters, even for the same probability distribution model, not to mention the parameters from different probability models. A then well-accepted solution was to find the common-*K* of the negative binomial distribution fitted to different data sets, hoping that the common-*K* to synthesize the information from multiple samples and to represent the species-specific characteristic of the biological species under investigation. Later, it was found that common-*K* is a dubious index for characterizing population spatial patterns and could not assume the role hoped for by entomologists (Taylor et al. 1979, Perry et al. 1986). In a series of studies (Taylor 1961, Taylor & Taylor 1977, Taylor et al. 1978, 1979, 1984, 1986), Taylor and his colleagues confirmed the wide applicability of Taylor's (1961) power law [Equation (1)] in describing the 'fat tails' of the distribution of abundance of nearly all organisms across multiple scales in both space and time; they also demonstrated the rarity of random distribution of biological populations. They further found that the parameter (exponent) *b* of Taylor's power law ranges from *1* to *2* for most species (hundreds of aphids, moths, and birds in Great Britain) they investigated.

Back to astronomy, Totsuji & Kihara (1969) obtained $r_0$=15.3x10$^6$ light years and *s*=1.8. According to Illian et al (2008), today the power law of the pair correlation function *g*(*r*) is generally accepted in astronomy (Martinez & Saar 2002), and modern estimates of the exponent (*s*) range from 1.7 to 1.8. The range of *s* [Equation (2)] is surprisingly similar to the range of *b*-value [Equation (10)] in biology (between 1 and 2) as Taylor et al discovered (Taylor 1986). The interpretation of the power law in astronomy is that there are no *characteristic scales* (Illian et al. 2008), which is a fundamental property of *scale-free network*. In a scale-free network, the degree distribution *P*(*k*) of network nodes (*k*) is found to manifestly departure from Poisson distribution predicted by random network theory and instead follows power law (Albert & Bararasi 1999, Newman & Barabasi et al. 2006, Barabasi 2009):

$$P(k) \propto k^{-\lambda}. \qquad (3)$$

It is well-known that the study of scale-free networks initiated by Albert & Bararasi (1999) and others in the 1990s is part of the still explosively growing network analysis studies in various disciplines towards a new discipline of network science (Albert & Bararasi 1999, Barabasi 2009). Obviously, the power laws Taylor (1961) and Totsuji & Kihara (1969) discovered in the 1960s are simply a manifestation of power law in the distributions of organism and galaxies, respectively. Nevertheless, I believe that there is a need to investigate the earlier discovered power laws (Taylor 1961, Totsuji & Kihara 1969) in the context of current network science since recent advances have produced some powerful mathematical methods and computational tools to analyze the mechanisms underlying power law relationships. It is noted that there is a practical difficulty to build network topology, as adopted in the current studies on scale free networks, with existing ecological data at the population level. But this difficult is largely gone at the community level, which explains the existing extensive studies on food web network in community ecology.

There are also parallel studies in astronomy and biology regarding some kind of threshold values of aggregation changes. In astronomy, $r_0$ is such a threshold value, which is equal to 15.3x10$^6$ light years and is a threshold value termed *minimum inter-point* or *hardcore distance*. In biology, Taylor (1981) proposed a Δ - model that tried to explain the mechanism of Taylor's power law, among other objectives. The model is with the following form:

$$\Delta = \varepsilon\left[\left(\frac{R_0}{R}\right)^s - \left(\frac{R_0}{R}\right)^t\right] \qquad (4)$$

where Δ is a displacement (movement) of an animal, *R* is the separation between animal individuals, $\varepsilon, s, t$ are parameters, $R_0$ is a threshold across which an equilibrium is reached. Ma (1988, 1991) studied the relationship between aggregation degree and population density and proposed the concept of population aggregation critical density (PACD). PACD can be computed with the following equation:

$$m_0 = \exp\left(\frac{\ln(a)}{(1-b)}\right) \qquad (5)$$

where *a* & *b* are parameters of Taylor's power law [Equation (1)]. Based on PACD and the notion that population aggregation is density-dependent, Ma (1988, 1991) reinterpreted Taylor's power law, which consists of a set of rules (which are based on the relationship among, population density, $m_0$, *a*, *b*, rather than b along as in the original Taylor's power law) for determining the spatial distribution patterns of biological populations.

In astronomy, power law is built with inter-point distance, and in entomology, power law is built with population density. Nevertheless, in entomology, distance has indeed been utilized in modeling insect dispersion patterns. It is generally agreed upon that insect dispersions generate distribution patterns that follow power law. For example, Taylor (1979, 1980) proposed a series of insect dispersion models, which relate the distance (*x*) from dispersion center with population density (*N*). Taylor found that the most general form of insect dispersion model was

$$N = \exp[a + bx^c + d\ln(x)], \qquad (6)$$

where *a*, *b*, *c*, & *d* are model parameters. I suggest that similar models to equation (6) should also be useful for studying human microbiome.

From above brief comparative review on the studies in both astronomy and biology, two points seem clear: (*i*) The power laws in astronomy and biology are isomorphic and they capture essentially the same or similar mechanisms underlying the distributions of galaxies and insects, respectively; (*ii*) Although in biology Taylor's power law has been applied only to the study of population distribution patterns, in the studies on the distribution of galaxies in the universe, there was not a similar limit, that is, power law has been applied to study the distribution of various galaxies, which are more similar to the community concept in biology. Therefore, it seems natural to investigate the applicability of Taylor's power law at the community scale in biology.

Furthermore, it is my opinion, there are at least three incentives to pursue the extending of power law for characterizing human microbial community: (*i*) The current surge of interests in studying human microbial communities initiated by US-NIH's HMP (Human Microbiome Project), fueled by the low-cost and high throughput new generation of sequencing technology, has generated unprecedented amount of data on human microbial communities. Given that each individual subject carries a microbiome consisting of multiple microbial communities (gut, skin, vaginal, oral, nasal, etc), even for a project that only investigates one type of microbial habitat (*e.g*., gut), there are multiple communities to be analyzed as long as there are repetitions of subjects (which are a must to conduct any statistical analysis) in a study. Then, one must compare and hopefully synthesize the community diversity information from many communities. Existing diversity indexes borrowed from macro ecology are weak or even lack the capability for synthesizing the information from multiple communities, except for simple statistical comparisons of the values of a diversity index. In addition, it is difficult to study the dynamics of community diversities with the existing diversity indexes. Power law can overcome these limitations of existing diversity indexes, as it has been demonstrated at the population level in biology and in astronomy. (*ii*) Taylor's power law essentially measures variability and heterogeneity, which exist at both population and community levels. Existing diversity indexes such as Shannon index, Simpson index, essentially measures the same properties (evenness and heterogeneity are simply reciprocal terms), but do not possess the capability to synthesize information from multiple communities. Therefore, even purely acting as a diversity index, power law should be superior to many of the existing diversity indexes. (*iii*) Power law parameters for the microbial community of a human individual have the potential to become a *personalized* (individualized) metric related to diseases and health, which can be invaluable for designing and implementing personalized medicine schemes for diseases associated with human microbial communities, since the values of power law parameters should be individual-dependent, perhaps more like finger printing. In other words, power-law approach to human microbial communities potentially offers a promising case study for medical-ecology-supported personalized medicine.

# A Proposal for the Extensions of Taylor's Power Law

Figure 2 shows a typical flow of metagenomic study of human microbiome. The product of the sequencing-bioinformatics pipeline is usually a species abundance table of gene reads in the format of Table (1) or Table (2) for cross-sectional experiments or longitudinal study, respectively. Table (1) contains the assumed gene reads data of 100 individual subjects from a cross-sectional study, *i.e.,* each individual subject is sampled only once. In Table (1), there would be abundance data of *S*=500 bacterial species from 100 samples, which represent 100 microbial communities. Table (2) contains the assumed gene reads data of 100 individuals from a longitudinal study, sampled from January 1$^{st}$ to December 1$^{st}$, say every 10 days. Then 36 samples for each individual subject would have been taken, and totally 3600 samples (100x36) or 3600 microbial communities to be analyzed.

In the study of human microbiome, it is also possible to get the abundance data at different locations in an organ, *e.g.*, population abundance of bacterial species at different locations of human vagina (*e.g.*, Kim et al. 2009). It is then possible to apply the dispersion models such as [Equation (6)] to describe the location-dependent abundance distribution. Table (3) shows an assumed data set with human vaginal microbial community as an example.

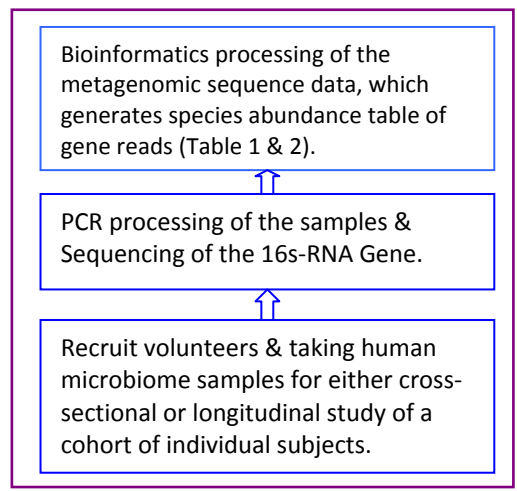

Figure 2. Pipeline of the metagenomic study of human microbiome.

Table 1. Assumed abundance table of 16sRNA gene reads of various bacterial species in human microbial communities generated from a typical cross-sectional study of 100 individual human subjects (numbered from #400 to #499).

| Subject Number | Species-1 Reads | Species-2 Reads | … … | Species-*N* N=500, Reads |
|---|---|---|---|---|
| #400 | 100 | 1200 | | 0 |
| #401 | 200 | 800 | | 1000 |
| … | … | … | | … |
| #499 | 1000 | 100 | | 50 |

Table 2. Assumed abundance table of 16sRNA gene reads of various bacterial species in human microbial communities generated from a typical longitudinal study of 100 individual human subjects (#400—#499) from January 1$^{st}$ to December 1$^{st}$.

| Subject Number | Date | Species-1 Reads | Species-2 Reads | … … | Species-N N=500,Reads |
|---|---|---|---|---|---|
| #400 | Jan 1 | 100 | 1200 | … | 0 |
| … | … | … | … | … | 1000 |
| … | Dec 1 | 500 | 2000 | … | 10 |
| #499 | Jan 1 | 200 | 1000 | … | 10 |
| … | … | … | … | … | … |
| … | Dec 1 | 700 | 900 | … | 0 |

Table 3. Assumed abundance data of abundance distribution at different locations of human vagina, with reference to a study conducted by Kim et al. (2009)

| Species | Upper 1/3 | Middle 1/3 | Lower 1/3 | Outer |
|---|---|---|---|---|
| Abundance of Species 1 | 100 | 900 | 1500 | 2000 |
| … | | | | |
| Abundance of Species n | 20 | 1600 | 100 | 70 |

With the assumed data formats displayed in Table (1) and (2), for cross-sectional and longitudinal studies, respectively, I envision five ways in which Taylor's power law models can be built to characterize the properties of human microbial communities such as community diversity and stability.

From a pure statistic perspective, building Taylor's power law model is a two-step process. The first step is to compute the pairs of mean [$M_i$] and variance [$V_i$] from the abundance data (which is in matrix form), and the second step is to fit the model, which can be as simple as performing a simple linear regression in the following form:

$$\ln(V) = \ln(a) + b\ln(M). \quad (7)$$

More complex algorithms to directly fit Taylor's power law [Equation (1)] are available in the literature [e.g., Marquardt's algorithm, see Ma (1990, 1992)], but the simple linear regression usually suffices.

It is noted that, when referring to the parameters of Taylor's power law, *b* is the most informative biologically and *a* is the least informative since *a* is mainly influenced by sampling schemes. Parameter $m_0$, *i.e.*, PACD (population aggregation critical density), offers insights on the dynamic relationship between aggregation degree and population density in population ecology, but its biological significance should be investigated further when the power law is extended to community scale.

## The Applications to Cross-sectional Studies

For a cross-sectional study, two ways for applying Taylor's power [Equation (1)] law should be possible: one approach is to compute pairs of the mean (*M*) and variance (*V*) across columns (one pair for each row) and another one is to compute *M* & *V* across rows (one pair for each column).

With the first way, the mean ($M_i$) and variance ($V_i$) reflect the average abundance and variability of all bacterial species detected in the communities, respectively. There is a pair of mean ($M_i$) and variance ($V_i$) for the community of each subject, and totally *S* pairs of *M* and *V*, that is, *i*=1,2,…*S*, where *S* is the number of subjects recruited in the study. The second step is simply to build a regression model with *S* pairs of means and variances based on Equation (7), and any statistical software would be able to perform the task. In the assumed case of Table (1), *S* would be 100 subjects, numbered from 400 to 499. The power law parameters obtained in this application manner would reflect the abundance variability of various bacterial species across *S* subjects. In other words, the parameters are determined by the variability of *N* bacterial species, and furthermore, the variability information among *S* subjects is also synthesized. To some extent, the parameters of Taylor's power law obtained in this type of application are similar to a "*common*" diversity index that synthesizes the diversity information of many subjects, similar to the common-*K* of the negative binomial distribution in the literature of the spatial distribution pattern of insect population (Taylor et al. 1979, Perry et al. 1986), but without the deficiency of the common-*K*.

In contrast, the second approach of applying Taylor's power law to cross-sectional study computes $M_j$ & $V_j$ across rows, where *j*=1, 2, … *N* species, *i.e.*, a pair of $M_j$ and $V_j$ for each bacterial species, and totally *N* pairs of $M_j$ and $V_j$. The follow step would be to obtain the parameters of Taylor's power law model by a simple linear regression analysis with the *N* pairs of data [Equation (7)]. The power law parameters obtained in this application manner reflect the variability of bacterial species abundance among individual subjects, as well as "synthesized" information across *N* bacterial species. To some extent, these power law parameters are similar to a "*common*" aggregation index of *N* bacterial species, since the term *aggregation* largely means *skewed variation* in the study of the spatial distribution of population abundance, and the term "*common*" again refers to some kind of "synthesis" of information from many (*N*) bacterial species, similar to the usage of term "common" in the common-*K* of negative binomial distribution as mentioned previously.

In summary, when Taylor's power law is applied to cross-sectional studies of human microbial communities, it may offer insights on "common" diversity measure of microbial communities for (across) multiple subjects, and "common" aggregation measure for (across) multiple bacterial species.

## The Applications to Longitudinal Studies

Obviously, cross-sectional data format is a special case of longitudinal data format, *i.e.,* a cross-sectional snapshot of the longitudinal data at a fixed time. Another way to examine the same data is to replace the time-series observations for each subject with their mean, and then the whole dataset of a longitudinal study is turned into a cross-sectional dataset. The above discussed two applications of Taylor's power law to cross-sectional data can be applied to the converted (taking the mean of time-series observations) longitudinal data in the same manner. However, as it will become clear from the following discussion, that the power law models built from longitudinal data, even after conversion, should reflect some dynamic properties of microbial communities. I classify this "mean-conversion" (*i.e.*, using the means of time-series observations) based application of Taylor's power law as the *third* approach to apply the power law for characterizing human microbial communities.

A next immediate question is, whether the longitudinal datasets can be used to fit Taylor's power law directly, without using the "mean-conversion" transformation. The answer should be a sound yes. First, the longitudinal data in the format of Table 2 can be used to fit Taylor's power law without combing time-series observations into mean values, by simply treat time-series observations as repetitions (repeated observations) of the subjects. I expect that the results from this approach may be similar to the above discussed "mean-conversion"-based analysis, but further investigation with real world metagenomic data should be performed to find out the difference, if there is any. Nevertheless, I believe that the most interesting and also the valuable applications of Taylor's power law to analyze longitudinal data should be performed on the individual-based as discussed below. In the following proposed application scheme, the dataset of each individual is used to build the power law model separately. This should be quite natural since all the time-series observations for each individual subject belong to the same microbial community. In other words, each individual carries a microbial community at a specific habitat (*e.g.*, gut). Therefore, from community ecology perspective, individual-based power-law modeling is truly at the community scale, and the previously discussed applications of the power law are more about meta-communities.

*Characterizing the stability of community diversity over time with Taylor's power law*

As stated previously, the dataset used for performing Taylor's power law analysis is individual based, that is, we apply the power law to characterize a single dynamic community, or various snapshots of the same community. For example, for the subject #400 in Table 2, we get an abundance dataset consisting of 12x3=36 rows (3 observations per month for 12 months) of the observations of $N$=500 bacterial species. Obviously, we can perform the same power law analysis to each of the subjects ($S$=100 in Table 2).

When Taylor's power law is applied to the dataset of an individual subject, there are also two ways that the application can be done, depending on how mean ($M$) and variance ($V$) are computed. In both kinds of applications, the computational procedures for building the power law model are actually the same as those used for the previously discussed modeling with cross-sectional data. Similar to the case of cross-sectional data, the computation of mean ($M$) and variance ($V$) can be cross-columns or cross-rows, corresponding to two ways Taylor's power law are applied for analyzing individual-based datasets.

For example, with the assumed datasets displayed in Table 2, for subject #400, crossing 500 columns (corresponding to $N$=500 bacterial species), we can compute 36 pairs (12 months, each month with 3 samples) of mean ($M$) and variance ($V$). The 36 pairs of mean-variance data can be used to fit Taylor's power law [Equation (7)]. Since each pair of the mean and variance reflects the heterogeneity (diversity) of the community at each time point, the parameters of the power law then should reflect the dynamic property of the community diversity over time series. In other words, the power law model built with cross time-series mean-variance pairs should reveal the stability of community diversity over time. I classify the Taylor's power built in this manner the *fourth* type of power law modeling of microbial community with a mission to reveal the stability of community diversity over time.

*Characterizing the stability of "common" species aggregation with Taylor's power law*

The mean/variance pairs can also be computed across rows for each subject. For example, for subject#400, we can compute one pair of mean and variance for each bacterial species across all of the 36 time-series observations; we can obtain as many as 500 pairs of the mean-variance values, *i.e.*, each species with one pair. The 500 pairs of mean-variance data can then be fitted to the power law [Eq. (7)]. Since each pair of the mean and variance computed across time series reflect the dynamic aggregation of each species, the power law model built in this way should represent aggregation dynamics common to all 500 species in the community. The term "common" species aggregation stability (*i.e.*, the stability of aggregation common to all species is obtained with this type of power law modeling.

In summary, when Taylor's power law is applied to longitudinal studies of human microbial communities, it may offer insights on the *stability of the community diversity over time series*, and on the *stability of "common" species aggregation*.

It is further noted that the power law modeling of human microbial communities when applied to longitudinal studies generates a set of power law parameters for each individual subject, which characterizes the dynamic properties of species aggregation and community diversity, or the stability of microbial community. This kind of insights can be invaluable for design and implementing medical-ecology-supported personalized medicine, as being argued in the next section.

**The Application of Dispersion Models**

The dispersion models such as Equation (6) proposed by Taylor (1979, 1980) can be used to describe location dependent population abundance data of bacterial species such as assumed in Table (3). It may be necessary to develop new dispersion models for human microbiome, other than directly applying Taylor's (1979, 1980) models, since unlike Taylor's power law, the existing dispersion models in literature are empirical and lack the necessary generality.

# PERSPECTIVES: A CASE STUDY FOR MEDICAL ECOLOGY AND PERSONALIZED MEDICINE

According to the website: www.medicalecology.org/, which is maintained by Dr. Dickson Despommier and Dr. Steven Chen of Columbia University, the term *Medical Ecology* was first coined by eminent microbiologist Rene Dubos, who believed that natural world could offer many of our needs if explored fully. Dubos was apparently inspired by events such as the discoveries of penicillin and gramicidin (his own discovery), in which soil microbes played a critical role, as well as the treatment of malaria with quinine. Recent redefinition of medical ecology has much broader meanings. For example, www.medicalecology.org/ stated that "*Medical Ecology is an emerging science that defines those aspects of the environment that have a direct bearing on human health. The concept of ecosystem functions and services helps to describe global processes that contribute to our well-being, helping to cleanse the air we breathe, the water we drink, and the food we eat. Environmental degradation often leads to alterations in these aspects, leading to various states of ill health.*"

In the post-HGP (Human Genomic Project) and HMP (Human Microbiome Project) era, it is high time to expand the scope of *medical ecology* again. It is my opinion that the study of the relationship between human microbiome and human host as well as its implications to human health and diseases should be considered as a core component of newly expanded *medical ecology*. I further believe that the emerging medical ecology is at the stage when *medical genetics* was emerging in the 1960s, and it should ultimately assume a similar role in medicine as today's medical genetics assumes. In the following discussion, I temporarily ignore the existing aspects of medical ecology; instead, I focus exclusively on the human-microbiome-centered *medical ecology*.

One may wonder if it is really necessary to redefine medical ecology rather than putting the related research topics in the context of some more familiar disciplines, such as *microbial ecology*, *clinical microbiology*, *bioinformatics*, and/or *eco-informatics*. In my opinion, no existing scientific disciplines can accommodate all of the interesting research themes that are waiting for medical ecology to embrace. Furthermore, to advance medical ecology, there will be some new theories and techniques to emerge, and these new theories and techniques need the umbrella of a redefined medical ecology. Figure 3 is a diagram showing the relationship between redefined medical ecology and its core parent fields, as well as its supporting and supported fields. In Figure 3, medical ecology is depicted as a cross-disciplinary and trans-disciplinary subject with "native parents" of medicine, ecology, and microbiology. Its advances are directly dependent on various *-omics* research (in particular metagenomics), computational biology, bioinformatics, system biology and system ecology, etc. It is my belief that medical ecology should become a foundation of personalized medicine, and that personalized medicine is likely to be the biggest beneficiary of research in medical ecology.

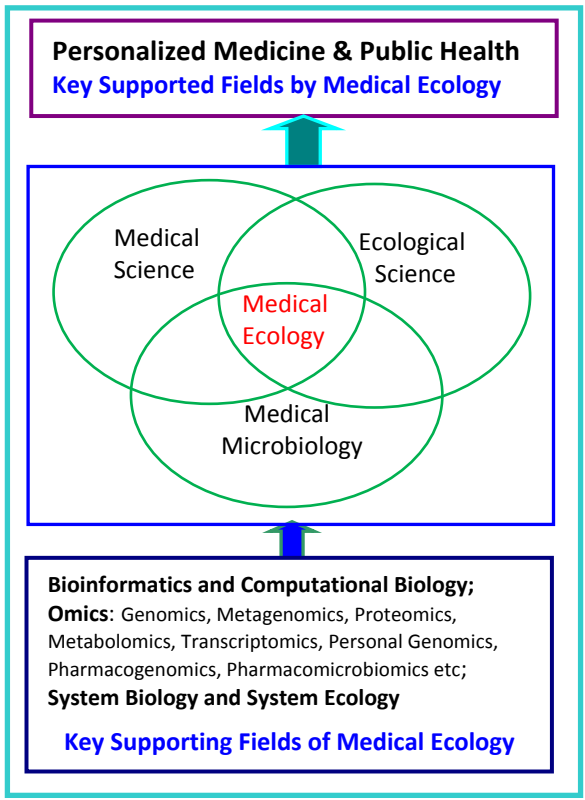

Figure 3. Emerging Human-Microbiome-Centered *Medical Ecology*, its parents, supporting and supported fields.

Research in medical ecology should develop some unique fields and theories in the near future. Obviously, predicting them is not an easy task, but here I try to identify three likely and important fields. It is widely recognized that ecology, evolution and behavior are three interwoven disciplines. Therefore, in Figure 4, I pick three research fields, each of which focuses on one of the three disciplines. These fields are: (*i*) applications and extensions of ecological theories from macro-ecology; (*ii*) evolution of symbiosis, cooperation and mutualism of "human microbiome/host" relationships, especially the role of human immune system; (*iii*) communication behavior of human microbiome and the communication between microbiome and host. Of course, this list of three is far from complete, and it is largely an impromptu attempt to present a glimpse of the research topics of emerging medical ecology. I will discuss the details of these topics in a separate paper.

Finally, I argue that the application of Taylor's power law for characterizing human microbial communities can serve as a case study of *medical ecology* and *personalized medicine* for the following reasons: (*i*) According to Figure 4, the study falls into the first category of the research topics of medical ecology, *i.e.*, the application of ecological theories, mostly from macro-ecology. Taylor's power law is one of a countable few laws in population ecology of animals and plants. (*ii*) As explained previously, the power law parameters, when obtained from longitudinal studies of human microbial communities, capture individual-specific and dynamic properties (community diversity and species aggregation) of an individual's microbial community. It should be possible to establish functional relationship between the individual-specific power law parameters and the metadata covariates that reflect the health/illness states of an individual. Understanding this kind of functional relationship can offer deep insights for practicing personalized medicine, especially for those diseases that are associated with human microbiome.

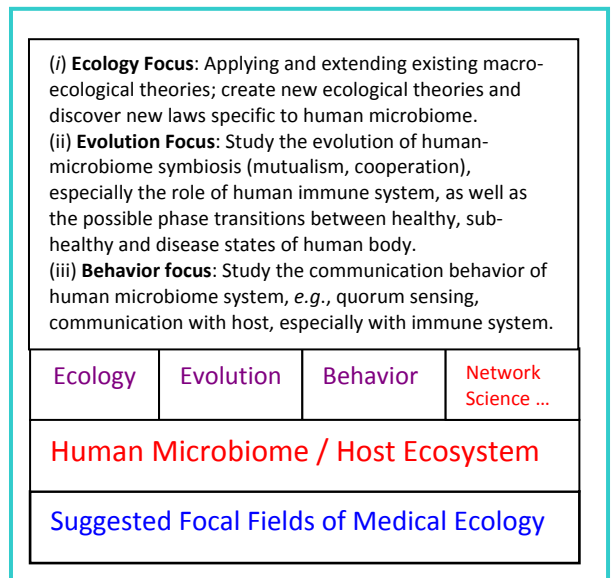

Figure 4. Suggested focal fields of Medical Ecology

In fact, scholars have already explored the application of Taylor's power law in epidemiology and evolutionary ecology of parasites (Morand & Krasnov 2008). For example, Morand & Krasnov (2008) suggested a possible link between host defense against parasite and the value of the parameter *b* of Taylor's power law (when applied to measure the aggregation of parasite populations). In personalized medicine, frailty analysis can be an effective statistical modeling approach (Ma et al. 2011), where the distribution of frailty (disease risk) can follow the PVF (power variance function) distributions. PVF is a family of probability distributions and is different from Taylor's power law. But in this distribution family, the *variance* and *mean* do satisfy power law function (which is the mathematical essence of Taylor's power law). It will be interesting to investigate the underlying biomedical process that may lead to the PVF frailty distribution. In a separate front, investigating the existence and maintenance of scale-free networks (of which power law is a fundamental property) in human microbiome should also be an important endeavor. Putting power law analysis in the context of network science (*e.g.*, Bascompte 2007, Ings 2009, Barabasi 2009) should produce synergetic advances in applying power law to the study of human microbiome and medical ecology.